\documentclass[preprint]{aastex}
\slugcomment{Accepted for Publication in the Astronomical Journal}
\shorttitle{Silicate Emission in TWA}
\shortauthors{Sitko, Lynch, and Russell}
\begin{document}
\title{Silicate Emission in the TW Hydrae Association}
\author{Michael L. Sitko\altaffilmark{1}}
\affil{ Department of Physics, University of Cincinnati, Cincinnati OH 45221-0011}
\affil{E-mail: sitko@physics.uc.edu}
\author{David K. Lynch\altaffilmark{1} and Ray W. Russell\altaffilmark{1}}
\affil{ The Aerospace Corporation, Los Angeles, CA 90009}
\altaffiltext{1}{Visiting Astronomer, NASA Infrared Telescope Facility, operated by the University of Hawaii under contract with the National Aeronautics and Space Administration.}

\begin{abstract}
The TW Hydrae Association is the nearest young stellar association. Among its members are HD 98800, HR 4796A, and TW Hydrae itself, the nearest known classical T Tauri star. We have observed these three stars spectroscopically between 3 and 13 $\mu$m. In TW Hya the spectrum shows a silicate emission feature that is similar to many other young stars with protostellar disks. The 11.2 $\mu$m feature indicative of significant amounts of crystalline olivine is not as strong as in some young stars and solar system comets. In HR 4796A, the thermal emission in the silicate feature is very weak, suggesting little in the way of (small silicate) grains near the star. The silicate band of HD 98800 (observed by us but also reported by \citet{ss96}) is intermediate in strength between TW Hya and HR 4796A.

\end{abstract}

\section{INTRODUCTION}

\begin{deluxetable}{cccccc}

\tablewidth{0pc}
\tablecaption{Observing Parameters}
\tablehead{
\colhead{Date (UT)} &
\colhead{Star}      & \colhead{Mean UT} &
\colhead{Mean Airmass} & \colhead{Number of Spectra} & 
\colhead{Int. Time (min)}}
\startdata

08 February 1998 & HD 98800  & 1135 & 1.46 & 4 & 32 \nl
08 February 1998 & Arcturus & 1205 & 1.45 & 3 & 4 \nl
21 March 1998 & Sirius & 0800 &  1.75  &  5 & 8  \nl
21 March 1998 & TW Hya & 0856 & 1.75 & 16 & 53 \nl
12 May 1998 & Pollux & 0751 & 2.78 & 5 & 7 \nl
12 May 1998 & HR 4796A & 0855 & 2.16 & 16 & 53 \nl
12 May 1998 & Vega & 1034 & 1.40 & 6 & 8 \nl

\enddata
\end{deluxetable}

Since the discovery 15 years ago that a large fraction of all main sequence stars possessed excess infrared emission \citep{aum88}, the study of these ``Vega-type stars'' and their precursors has grown enormously. With the detection of the edge-on debris disk in $\beta$ Pictoris \citep{st84}, and the equally important discovery that the debris exhibited a silicate emission band whose structure was similar to solar system comets \citep{knacke93} it was apparent that at least some of these systems might be true planetary systems, perhaps similar to what our own solar system was like in its youth. Since that time, disks and tori around other stars have been successfully imaged.

Recently, \citet{kastner97} identified a previously unrecognized nearby young stellar association that consisted of 5 T Tauri (TT) or T Tauri-like stars, including the classical TT star TW Hya. Designated as the TW Hya Association (TWA), the list of objects in this stellar group is now over a dozen objects and growing \citep{webb00}. Located at a distance of about 50 pc, TWA represents the nearest region of recent ($<$20 Myr) star formation. It is no longer embedded in a molecular cloud, yet one of its members, TW Hya itself, possesses both a circumstellar molecular cloud (CO, HCN, and CN have been detected; \citet{kastner97}) and an infrared excess. HR 4796A, which is also believed to be a member of TWA \citep{webb00} has a dusty torus that has been recently resolved \citep{rayjay98,koerner98,schneider99}. Another member of TWA is HD 98800, a multiple star system with dust concentrated around one unresolved pair \citep{gehrz99,koerner00}. Combining a variety of criteria, \citet{weintraub00} derive an age of 5-15 Myr for TWA. 

Because TWA is the nearest group of pre-main sequence stars known, it is providing imaging possibilities that are in some ways superior to that of the Taurus-Auriga complex, and at an epoch of great importance to the study of planet formation. This association provides a laboratory for studying a number of stars of the same age but different masses, multiplicities, etc.  Although the luminosities of low-mass stars in TWA are substantially less than those of younger stars of the same mass (such as in Taurus-Auriga), the relative proximity of the TW Hya association and the implementation of sensitive IR array detectors on large telescopes opens up a realm of study of PMS disk evolution not possible before.

In this paper we report observations of the 3-13 $\mu$m spectra of TW Hya, HR 4796A, and HD 98800. The mid-IR spectrum of HD 98800 has already been reported by \citet{ss96} using ground-based data, and by \citet{wh00} using ISOPHOT. Our data extends to shorter wavelengths than the Sylvester \& Skinner data, while having a better signal-to-noise ratio than the ISOPHOT observations. Both are useful for realistically determining the photospheric contribution to the observed emission.
 
\section{OBSERVATIONS}

Spectra of the TWA targets were obtained with the Aerospace Corporation's Broadband Array Spectrograph System (BASS) configured to f/35 at the 3m NASA Infrared Telescope Facility (IRTF). The BASS consists of a pair of cooled prisms which disperse the spectrum onto two 58-element blocked impurity band (BIB) linear arrays that simultaneously span the 3-13.5 $\mu$m region. The spectral dispersion ranges from about 30 to 125 over each of the 3-6 and 6.5-13.5 $\mu$m regions.  At the IRTF, the BASS entrance aperture subtends 3.4 arcsec. The observations of TW Hya were obtained on 21 March 1998 (UT). HD 98800 was observed on 8 February 1998 (UT). HR 4796A was observed on 12 May 1998 (UT).

\begin{figure}
\plotone{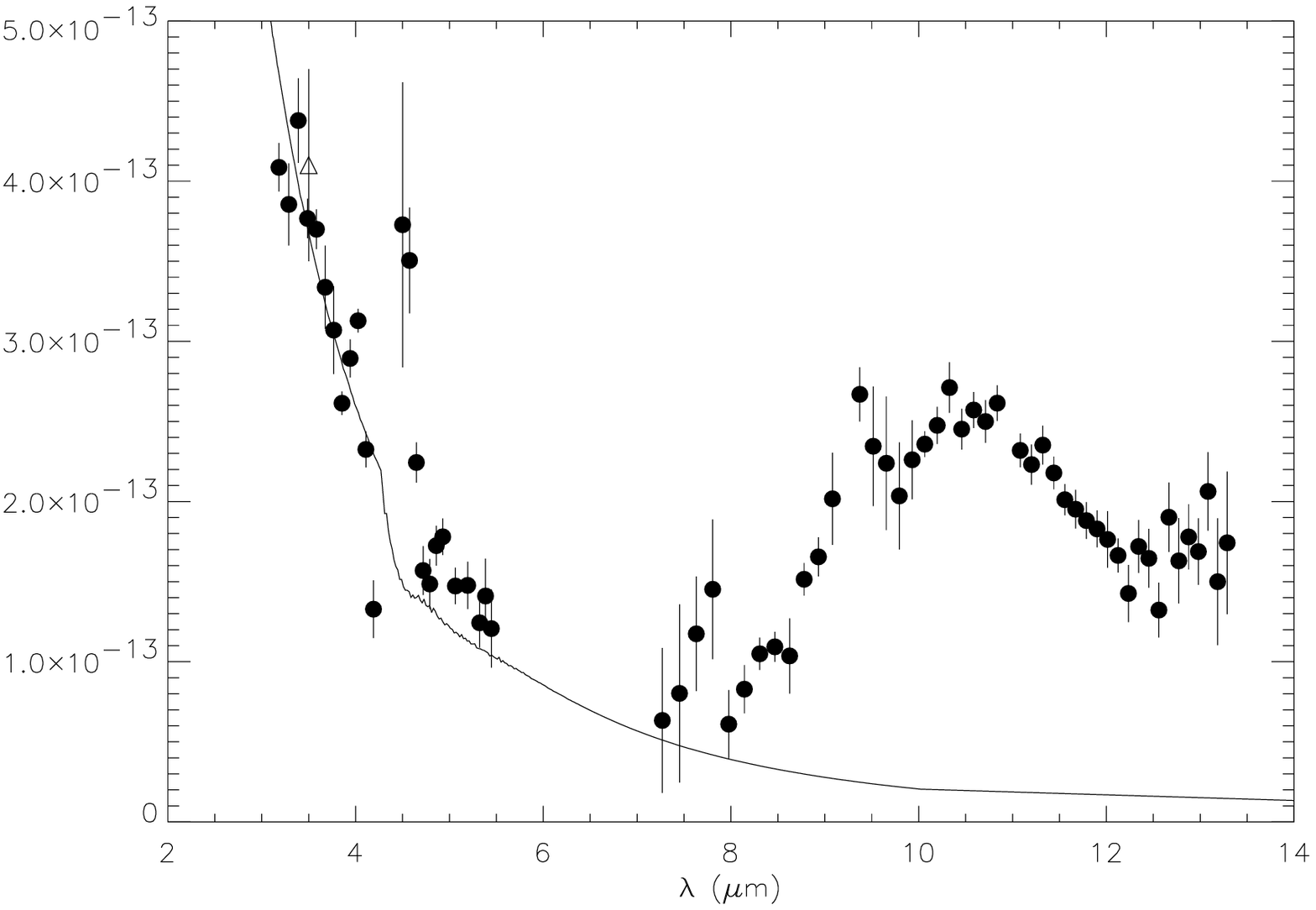}
\figcaption[Sitko.Fig1.eps]{The spectrum of TW Hya (filled circles) from 3-14$\mu$m, along with the L-band data of \cite{rk83} (open triangle, barely visible amongst the BASS data), and the IRAS 12 $\mu$m data. No adjustments to the flux levels of any of the data have been made. For comparison, we also show the K7 model atmosphere (T=4000 K, log g=4.5, log z=0.0). The model spectrum was generated using the 1991 Kurucz models, reproduced using the IUEDAC's KURUCZ91.PRO routine, and normalized to the L-band data of Rucinski and Krautter. \label{fig1}}
\end{figure}

\begin{figure}
\plotone{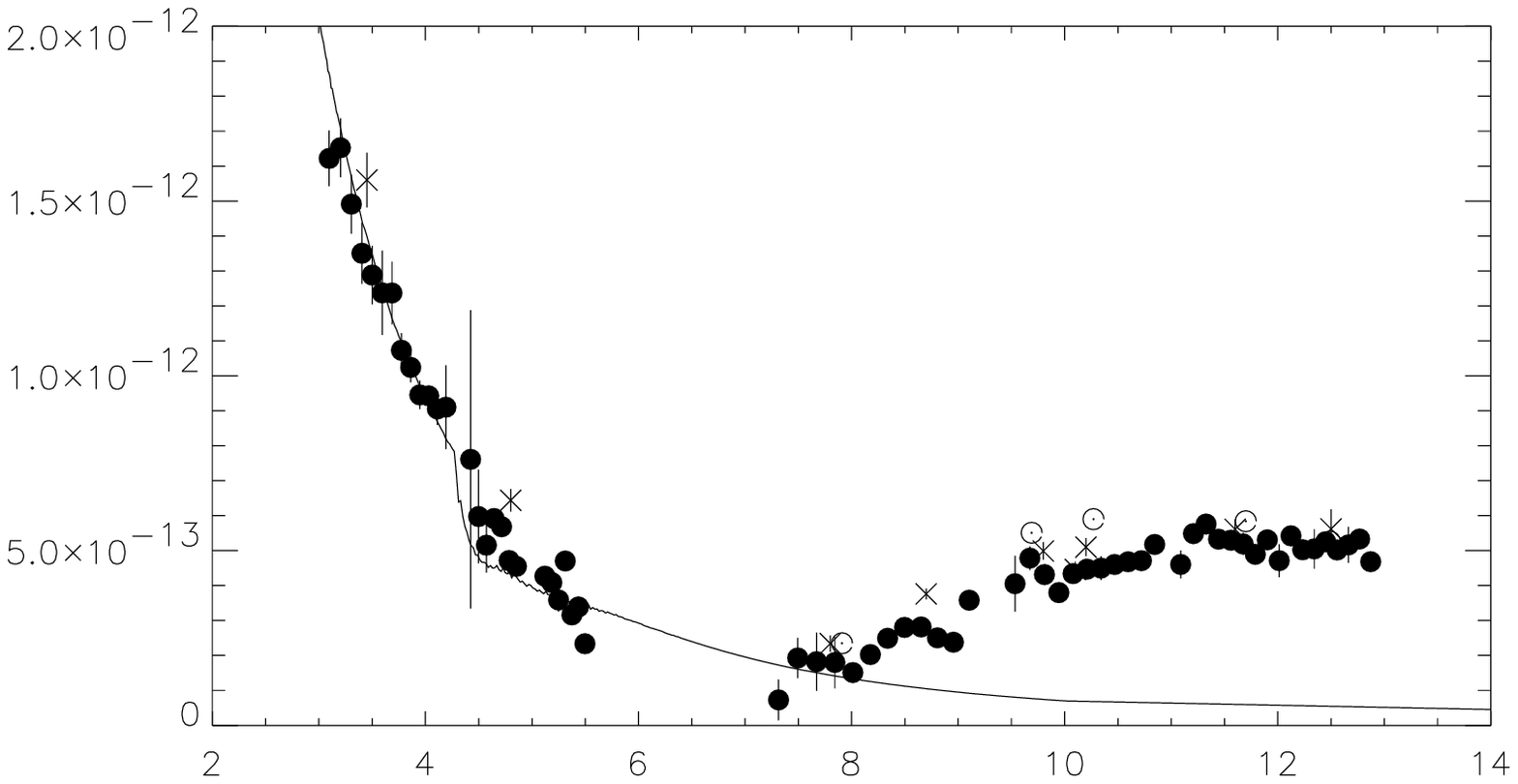}
\figcaption[Sitko.Fig2.eps]{The spectrum of HD 98800 (filled circles) from 3-14 $\mu$m. Also shown is the appropriate model atmosphere (T=4250 K, log g=4.5, log z=0.0), normalized to the flux at 3.5 $\mu$m. For comparison, we have also plotted the photometric data of \citet{zb93} as 'x' symbols and \citet{koerner00} as open circles. \label{fig2}}
\end{figure}

Details of the observations are listed in Table 1, and the resultant spectra are shown in Figures 1 (TW Hya), 2 (HD 98800), and 3 (HR 4796A), where we have plotted the spectral flux ($\lambda$F$_{\lambda}$ in W m$^{-2}$). All BASS spectral data points such as those in the atmospheric CO$_2$ and H$_2$O bands with errors exceeding the observed flux (i.e., signal/noise less than unity) have been rejected. The spectrum of TW Hya was flux-calibrated using observations of $\alpha$ CMa (Sirius) at approximately the same airmass, obtained just prior to the TW Hya data. The spectrum of HD 98800 was calibrated using observations of $\alpha$ Boo (Arcturus) at the same airmass, and converted to the same flux system using the Arcturus/Sirius flux ratio determined by \citet{rm98} for the BASS instrument. For HR 4796A, no calibrator at the same airmass were observed. To establish the spectrum, it was calibrated against one star, $\alpha$ Lyr (Vega) at a lower airmass, and another star, $\beta$ Gem (Pollux) at higher airmass. For the latter, we also used the Pollux/Sirius ratio of \citet{rm98}. Outside of those wavelengths that are severely affected by telluric lines, and which are usually rejected from the figures due to their inferior signal-to-noise ratio (SNR), the resultant fluxes are nearly identical. We illustrate this fact by plotting both resultant spectra in Figure 3. Because of the similarity of these spectra, we have taken the mean values and plotted them against the appropriate model atmosphere spectrum in Figure 4. The detail near the 10 $\mu$m band is shown in Figure 5.

\begin{figure}
\plotone{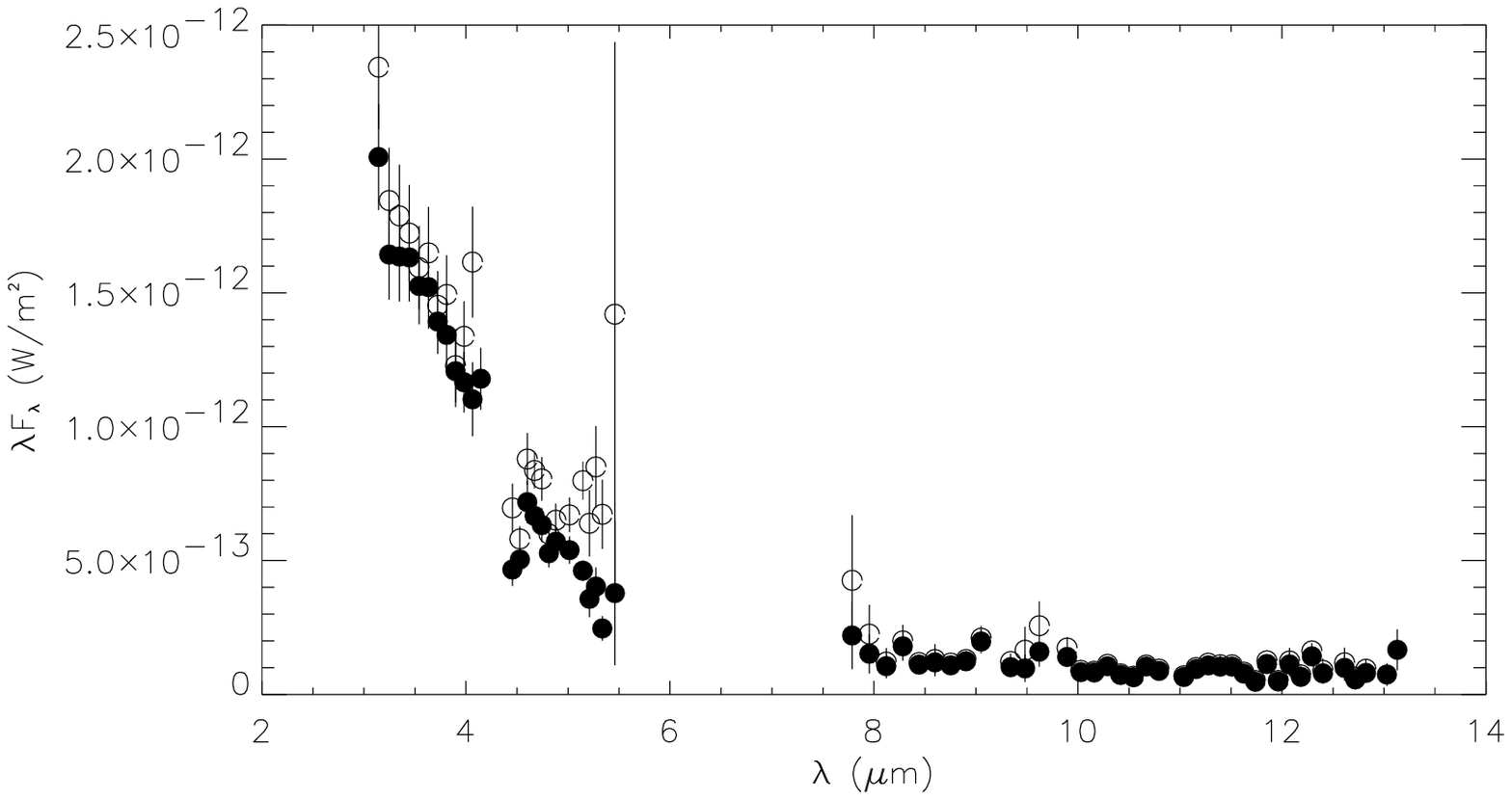}
\figcaption[Sitko.Fig3.eps]{The spectrum of HR 4796A from 3-14 $\mu$m, calibrated using $\alpha$ Lyr (filled circles) and $\beta$ Gem (open circles). \label{fig3}}
\end{figure}

\begin{figure}
\plotone{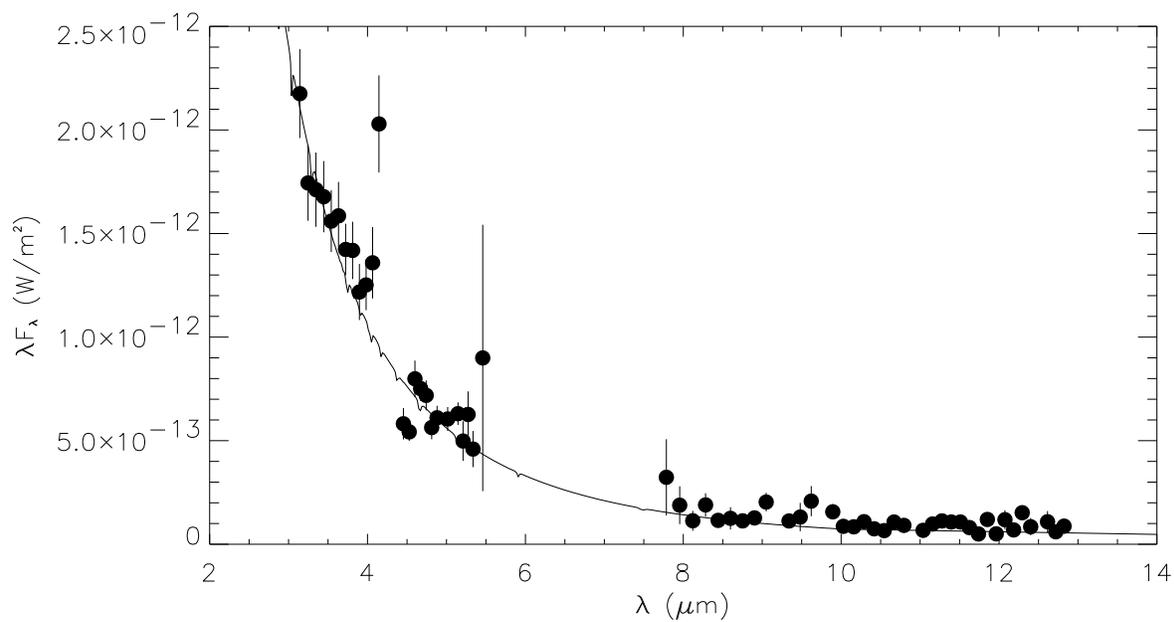}
\figcaption[Sitko.Fig14eps]{The spectrum of HR 4796A, and the appropriate model atmosphere (T=9500 K, log g=4.0, log z=0.0), normalized to the flux at 3.5 $\mu$m. \label{fig4}}
\end{figure}

\begin{figure}
\plotone{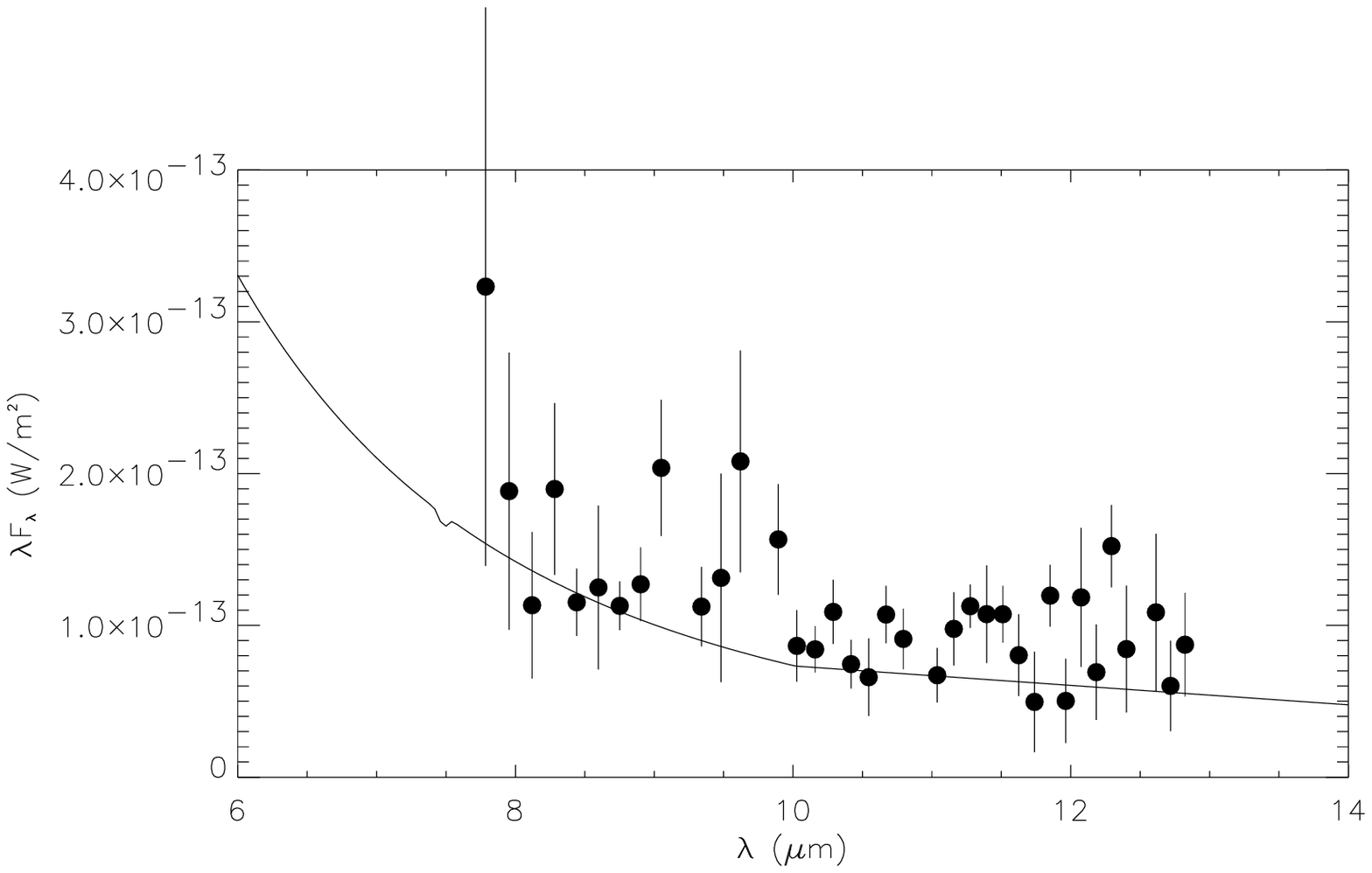}
\figcaption[Sitko.Fig5.eps]{The spectrum of HR 4796A, detail near 10 $\mu$m. \label{fig5}}
\end{figure}

\section{Discussion of the Spectra of Individual Objects}

\subsection{TW Hya}

Figure 6 shows the spectral energy distribution of TW Hya from 0.44 to 100 $\mu$m. The photometric data in the 12, 25, 60, and 100$\mu$m bands were obtained by IRAS. In addition, measurements in the B, V, R, I, J, H, K, and L photometric bands by \cite[]{rk83} are included. Because TW Hya is a variable star, we show the mean value of the Rucinski and Krautter measurements, while the error bars indicate the range in those values. We also show the Kurucz model photosphere spectrum of a K7 V star, normalized to the K and L-band flux. 

\begin{figure}
\plotone{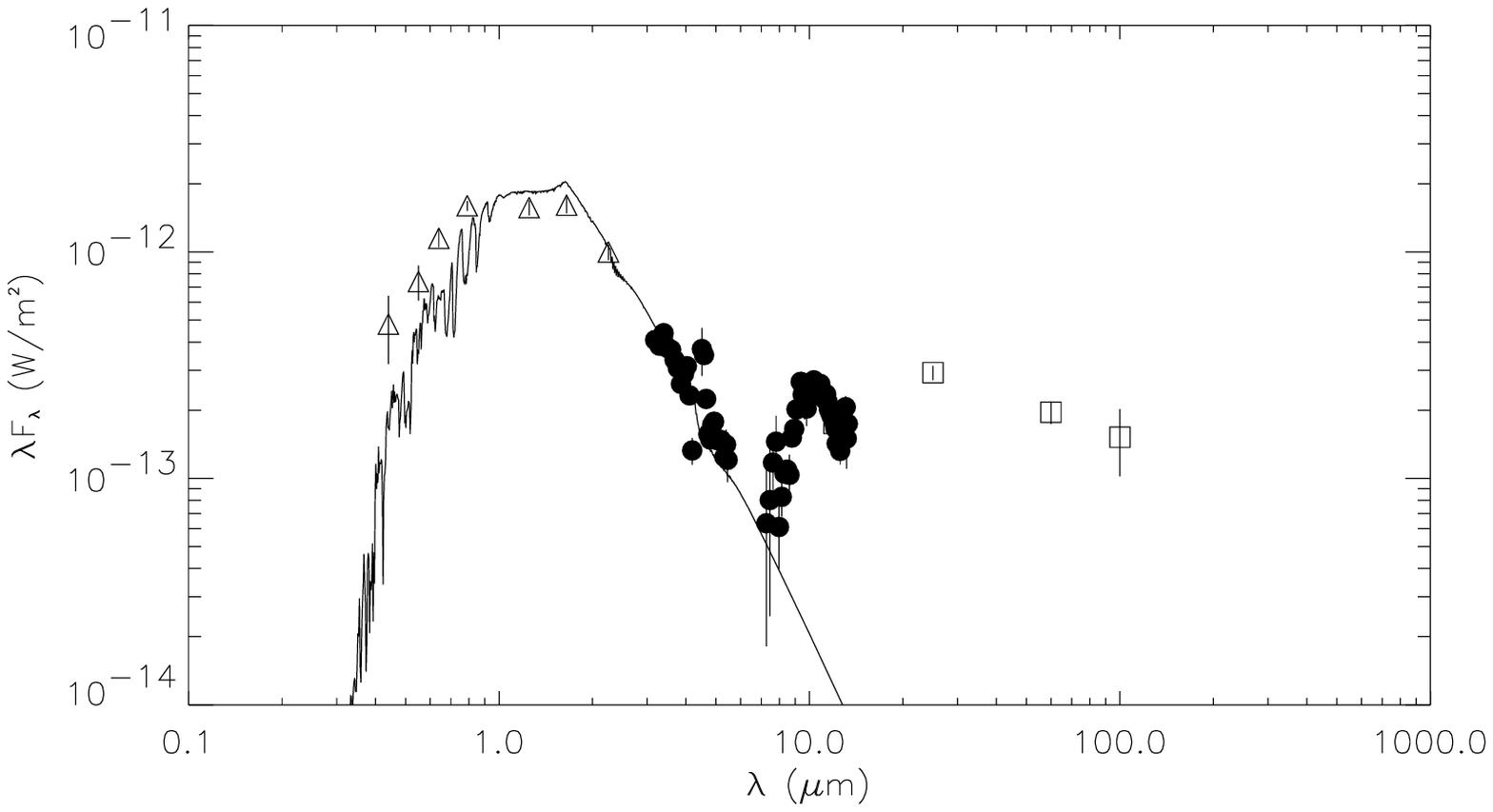}
\figcaption[Sitko.Fig6.eps]{The total energy distribution of TW Hya from 0.44 $\mu$m to 100 $\mu$m. The filled circles are data obtained with BASS. The open triangles are the data of \cite[]{rk83}, and the open squares are the IRAS photomtery. \label{fig6}}
\end{figure}

As can be seen in Figure 6, TW Hya possesses a significant infrared excess. In fact, the emission longward of 8 $\mu$m, which is dominated by the thermal emission of the circumstellar dust, accounts for 23\% of the total observed luminosity of the star. Originally, \citet{rk83} fit the data at 1.25 $\mu$m and shortward with a blackbody spectrum, which left a flux excess between 1.25 and 3.5 $\mu$m. However, model atmospheres of stars in this temperature range have line opacities that naturally reproduce the shape of the spectrum observed in this region. In Figure 6 we illustrate this by fitting a Kurucz model atmosphere of a K7 star, and normalize it to the 2-4 $\mu$m BASS data. The same model fit is included in Figure 1. The fit is not perfect down to the shortest wavelength, but TW Hya is a variable star, the BASS data are not simultaneous with the observations of Rucinski and Krautter, and the star possesses a short-wavelength excess attributed to the disk-star boundary layer \citep{muzerolle00} which we have not included here. In any case, there does not appear to be any significant excess flux between 2 and 8 $\mu$m.

In TW Hya (see Figure 1) the emission between 8 and 13 $\mu$m is dominated by the silicate band at 10 $\mu$m. The three points between 9.5 and 9.8 $\mu$m with the large error bars lie in the telluric ozone band and may not be reliable, even though the data on TW Hya were obtained at the same airmass as that of Sirius. The point at 9.4 $\mu$m with the smaller error bar sits right at the edge of this band.

\subsection{HD 98800}

The spectrum of HD 98800 between 3 and 13.5 $\mu$m is shown in Figure 2. Also shown in the figure are the photometric data of \citet{zb93} and \citet{koerner00}. The agreement between absolute fluxes derived from the BASS data and the other two sets is good, although the BASS fluxes are slightly lower. The shape and total flux level is also the same as the spectral data obtained by \citet{ss96}, whose observations cover the 8-24 $\mu$m region, and who used the photometry from another instrument in their modeling. The value of the BASS observations are that they cover the wavelength regions dominated by both photospheric and dust emission simultaneously and on the same photometric system, allowing the photospheric flux to be extrapolated out into the 10 $\mu$m region using a single set of simultaneous data. The ISOPHOT observations of \citet{wh00} cover the 5.8-11.6 $\mu$m region, but suffer from such large systematic fluctuations in the signal that the silicate feature cannot even be seen.

The BASS spectrum shortward of 6 $\mu$m is well-fit by the photospheric model normalized near 3 $\mu$m. This is consistent with the results of \citet{koerner00}, who fit similar models to the individual A and B components of the system (each of which is a binary), and find that the flux is dominated by photospheric emission out to 7 $\mu$m in both A and B components.

\subsection{HR 4796A}

In Figures 4 and 5 we show the spectrum of HR 4796A and its appropriate model atmosphere. The 8-13.5 $\mu$m flux is weak, and half is photospheric. \citet{koerner98}, using thermal imaging with the MERLIN instrument on Keck, derive a total flux at 12.5 $\mu$m of 5.4x10$^{-14}$ Wm$^{-2}$, of which it was estimated that half was photospheric. For the BASS observations, the 12-13 $\mu$m flux is higher than those derived from MERLIN, but the fraction due to photospheric emission is comparable. The lack of any detectable excess flux shortward of 8 $\mu$m indicates that there is little dust hotter than 450 K or closer than about 2 AU (depending somewhat on the optical albedo and infrared emissivity of the grains). \citet{aug99} have  modeled the spectral energy distribution of HR 4796A, including the hot inner region, with a 2-component dust model. In their model, the excess emission near 10 $\mu$m comes primarily from huge (larger than 100 $\mu$m in size) grains located near 9 AU from the star. Unfortunately, the quality of the spectra here are inadequate the detailed spectral features expected from such material.

\section{COMPARISON WITH SOLAR SYSTEM COMETS AND OTHER DUSTY PMS STARS}

\subsection{General Considerations}

Some main sequence and pre-main sequence stars with dusty debris disks possess silicate grains whose spectral features resemble those of long-period comets. \citet{knacke93} were the first to demonstrate this by comparing the 10 $\mu$m emission feature in $\beta$ Pic with 1P/Halley and Levy 1990 (C/1991 L3). All three objects exhibit an emission band with maxima or shoulders near 9.5 $\mu$m and 11.2 $\mu$m, the latter indicative of crystalline olivine. Subsequently, HAEBEs embedded in star-forming regions have been examined (i.e. \citet{hanner95}) and some, such as HD 150193 in the $\rho$ Oph cloud, are found to possess features of a similar nature. The two isolated HAEBEs (which might also be called post-HAEBEs since they no longer embedded in nebulosity) HD 31648 and HD 163296, were shown to have spectra similar to that of Hale-Bopp and Levy \citep{sitko99}. However, the stellar disk dust in these systems emitted relatively more strongly near 9.4 $\mu$m than the comets.

It should be remembered that in the case of a comet, we are dealing with material with a single well-defined distance from the sun, while this is not the case for a disk. Nevertheless, \citet{wooden00} have shown that over a range of a factor of three in heliocentric distance, the spectral shape of Hale-Bopp was, with the exception of one subfeature, invariant. Furthermore, a comparison of the ISO spectra of the HAEBE star HD 100546 and Hale-Bopp indicate that, while the overall spectral energy distributions are significantly different in the way expected (more cool dust in the stellar disk), the shapes of the narrow features were still quite similar \citep{malfait98}. However, the shape of the 10 $\mu$m band is sensitive to the abundance and form of pyroxenes, as evidenced in the variety of spectral shapes of pyroxene-rich interplanetary dust particles (IDPs) that have been analyzed in the laboratory \citep{sw85,brad92}. It is interesting that the one subfeature that {\it was} observed to vary with heliocentric distance in Hale-Bopp has been attributed to crystalline pyroxenes \citep{wooden00}. This suggests that perhaps the main reason for the excess 9.4 $\mu$m emission of the disks compared to comets is due to the presence of dust at temperatures much higher than those generally observed in solar system comets, which allows the crystalline pyroxenes to be enhanced, but on a grander scale than was seen in Hale-Bopp.

\subsection{TW Hya}

In Figure 7 we show the 10 $\mu$m band of TW Hya, normalized to the continuum by dividing by a gray body that matches the observed flux at 8 and 13 $\mu$m. Also shown are the data for HD 31648, HD 163296, comet Hale-Bopp, and comet Levy from \citet{sitko99}. Although the data for TW Hya has a modest SNR, it is apparent that TW Hya does not have the same shape as the other objects. In particular, the well-defined crystalline olivine feature seen at 11.2 $\mu$m in both the comets and HD 31648 and HD 163296 seems to considerably less-defined in TW Hya. At best, a modest change of spectral slope is present at that wavelength. In Figure 8 we compare the spectrum of TW Hya with three HAEBEs \citep{sitko00} whose emission features are representative of the wide variety that are seen in these PMS objects (excluding objects dominated by organic emission features). That of RY Tau exhibits a silicate feature which is strong and relatively featureless, except for a single maximum near 9.7 $\mu$m. In HD 163296 (again) the olivine feature is present, but not as prominent as in comets such as Hale-Bopp and Levy. Finally, the spectrum of HD 35187 has a weak silicate band, but more comet-like (flat-topped) in shape. In Figure 9 we also compare the spectrum of TW Hya to that of RW Aur, a PMS G5e star with circumstellar CO emission in the first vibrational overtone band \citep{sh00}. Because the strength of the silicate band with respect to the underlying continuum in RW Aur is only one-quarter that of TW Hya, for comparison purposes the band strength with respect to the continuum in RW Aur has been scaled up by a factor of 4.

Of the four ``comparison'' star, the silicate band in TW Hya is most similar {\it in shape and strength} to that of RY Tau, a F8Ve PMS star in the Taurus-Auriga cloud. The spectral {\it shape} in the silicate band of TW Hya is also nearly identical to that of RW Aur (G5e), but its strength is considerably stronger. This should not be construed to imply that TW Hya and the other late-type PMS stars are systematically different from the earlier-type HAEBEs. Most HAEBEs are not strong 11.2 $\mu$m emitters. Furthermore, high-quality mid-IR spectra on a large sample of TTs are lacking.

\begin{figure}
\plotone{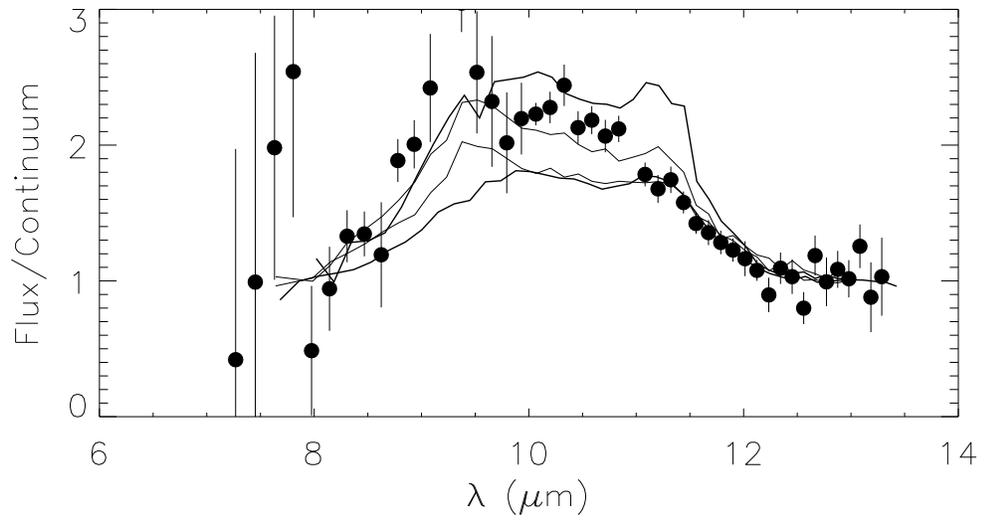}
\figcaption[Sitko.Fig7.eps]{The flux/continuum ratio for TW Hya (filled circles) compared to those of HD 163296 (upper thin line), HD 31648 (lower thin line), Comet Hale-Bopp (upper thick line) and Comet Levy (lower thick line). \label{fig7}}
\end{figure}

\begin{figure}
\plotone{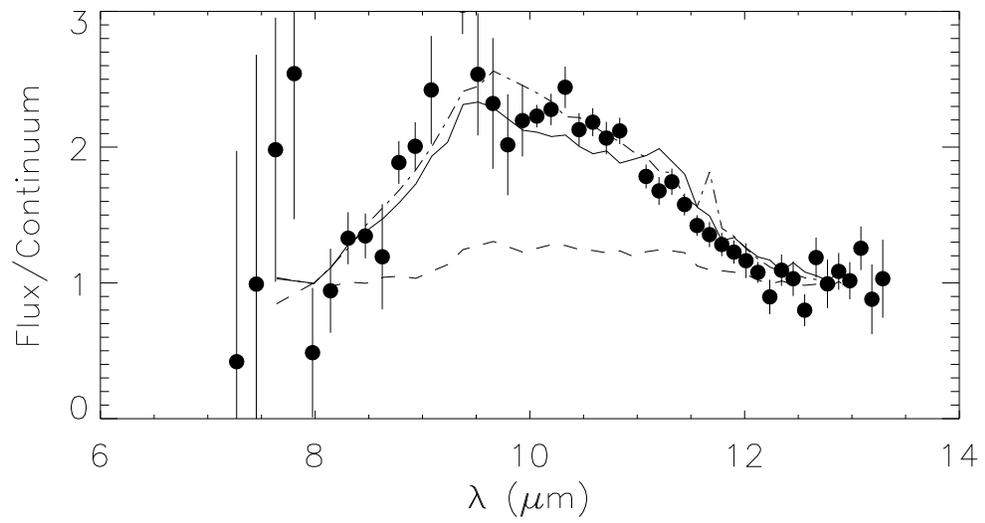}
\figcaption[Sitko.Fig8.eps]{The flux/continuum ratio for TW Hya (filled circles) compared to that of HAEBEs with a wide variety of spectral shapes: HD 163296 (solid line), HD 35187 (dashed line) and RY Tau (dot-dashed line).\label{fig8}}
\end{figure}

\begin{figure}
\plotone{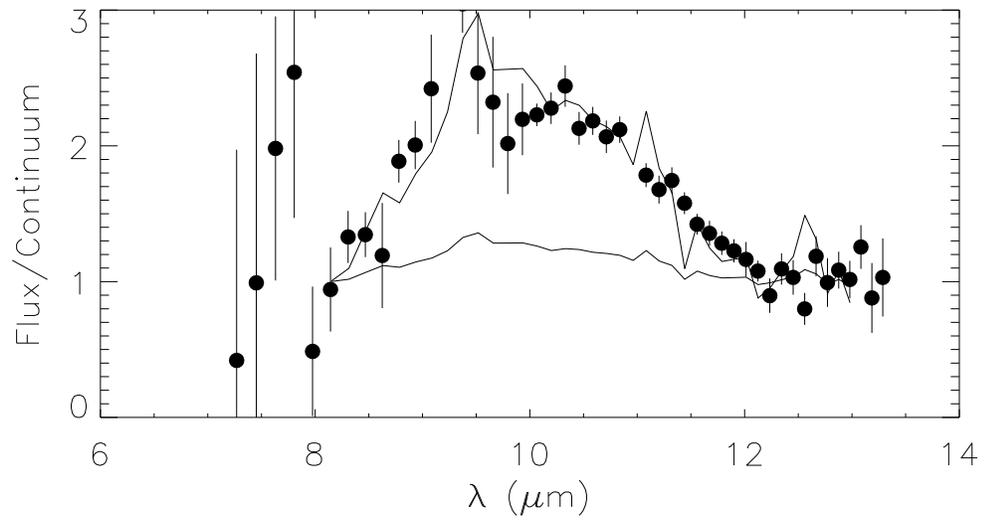}
\figcaption[Sitko.Fig9.eps]{The flux/continuum ratio of TW Hya (filled circles) compared to that of RW Aur (solid line). The flux/continuum ratio of RW Aur has been scaled by a factor of four in this figure. \label{fig9}}
\end{figure}

\begin{figure}
\plotone{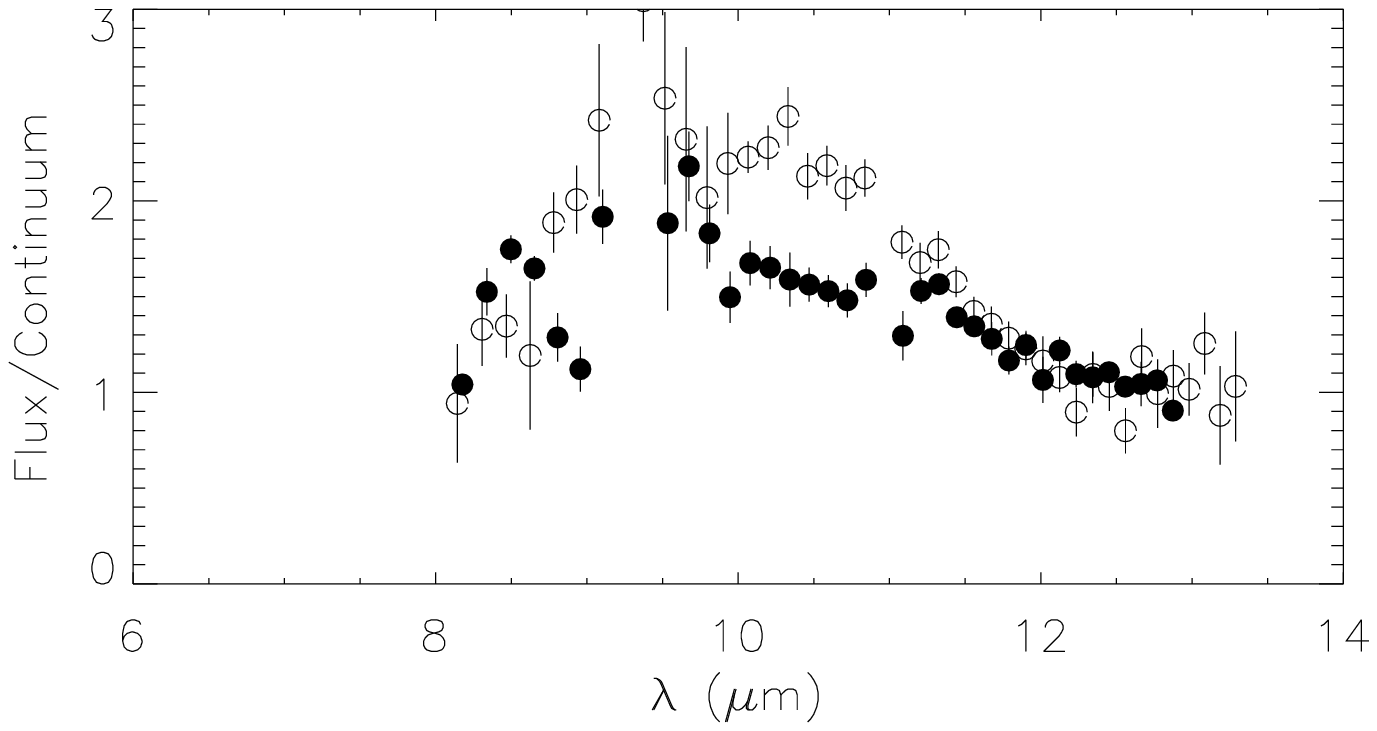}
\figcaption[Sitko.Fig10.eps]{The flux/continuum ratio of HD 98800 (filled circles) compared to that of TW Hya (open circles). \label{fig10}}
\end{figure}
 
In Figure 10 we show the flux/continuum ratio of HD 98800 with that of TW Hya. While the two sets of data overlap reasonably well between 11 and 13 $\mu$m, much of this may simply be due to the normalization procedure used to get the flux/continuum ratio. Overall, the band strength seems to be weaker in HD 98800, and there may be structural differences as well. However, when the band strength is intrinsically weak, the normalization procedure tends to magnify the systematic effects in the data (such as small errors in cancellation of telluric lines between target and calibration star) so that spectral ``features'' may not be real. In such cases, a direct comparison of the observed fluxes may be more useful. In Figure 11 we compare the observed fluxes of these two stars directly (after suitable re-scaling). Here, the differences are more readily apparent.

\begin{figure}
\plotone{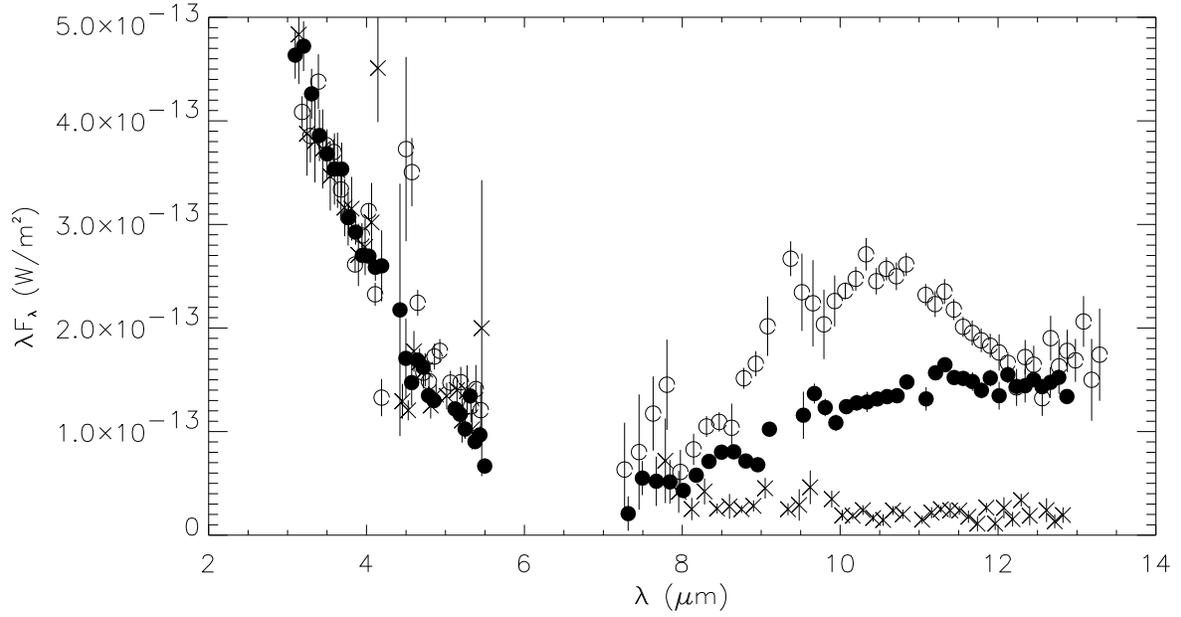}
\figcaption[Sitko.Fig11.eps]{The observed flux of HD 98800 (filled circles), TW Hya (open circles), and HR 4796A ('x' symbols) compared directly. In this figure, the flux of HD 98800 has been scaled down by a factor of 3.5, and that of HR 4796A by a factor of 4.5. \label{fig11}}
\end{figure}

Due to the weakness of the dust emission in HR 4796A compared to the extrapolated photospheric flux, the flux/continuum ratio was not determined.

\section{DISCUSSION}

In Figure 11 we compare the spectra of all three stars, normalized to the same flux between 3 and 5 $\mu$m, where all three are dominated by photospheric emission. It is apparent that the net strength of the silicate emission feature of the dust surrounding the three TWA stars is different, and it cannot simply be the result of differing ages, since the stars are believed to be approximately coeval. Thus the degree to which small silicate grains survive and/or are modified in these systems must be dependent on other parameters as well, such as stellar mass, presence and location of companions, and the like. With only three TWA stars observed so far, it is premature to draw any specific conclusions in this regard. Furthermore, the quality of the spectra of the dust in HR 4796A and HD 98800 preclude any sweeping generalizations concerning their detailed mineralogies. Instead, we examine some of the inferences that can be made concerning the general nature of the material surrounding TW Hya specifically.

The 11.2 $\mu$m crystalline olivine feature has not been detected with the ISO Short Wavelength Spectrometer (SWS) in the emission spectra of the youngest (i.e. embedded) PMS stars, but is seen observed in a number of more evolved objects \citep{waelkens00}. Because the majority of the objects observable with ISO SWS must be bright, most of the targets have been HAEBEs. The TTs were too faint for SWS \citep{mario00}. Only a few were observed at more modest spectral resolution with ISOPHOT (see \citet{natta00}), but much of these data have inadequate signal-to-noise to make any conclusions regarding the presence of crystalline material.

 The dust in TW Hya itself does not appear to exhibit the degree of crystallinity, as judged by the strength of the 11.2 $\mu$m olivine feature, that some other objects with resolved debris disks have, and which is seen in many, {\it but not all}, solar system comets \citep{hanner94}. For example, in the HAEBE star HD 163296, a star whose age is believed to be between 4 and 6 Myr, the feature is definitely present \citep{sitko99,vda00}. About half (or more) of long-period comets studied so far also possess the 11.2 $\mu$m feature. These comets have their recent origin in the Oort cloud, and were presumably originally formed in the Uranus-Neptune region of the early solar system. By contrast, short-period comets, whose origin is probably in the Kuiper belt, tend to have weaker silicate features \citep{lynch92,lynch95,lynch00,hanner96}. The mean particle sizes must be greater (or the feature otherwise hidden in some fashion) in these latter objects, and the degree of crystallinity cannot be assessed. 

When does the crystalline material form in these disk systems, if at all? \citet{brad99} have demonstrated that the most primitive unequilibrated interplanetary dust particles, of presumed cometary origin, exhibit a smooth silicate band essentially devoid of crystalline features. This is consistent with their glassy internal structure. The spectra of these particles resemble those of disk systems that do not exhibit the 11.2 $\mu$m feature. However, the large porous grains in which they are embedded, {\it along with significant amounts of crystalline material}, have spectra which resemble those older disk systems and solar system comets that do exhibit this crystalline feature. It may be that much of the crystalline material is produced early in the development of the protosolar nebula, and is later agglomerated with the more pristine material \citep{brad00}. One could even envision a scenario whereby disk or comet dust shed and heated near the star is recycled over the entire lifetime of the formation of the bulk of the cometesimals in the nebula. In such a scenario, ``pristine'' objects might be highly ``contaminated'' by dust processed at a variety of temperatures.

The lack of the feature in TW Hya, whose age is greater than that of some other stars exhibiting the feature, could be due to a number of factors. Perhaps it is simply too young, and hasn't entered the stage where this material is produced. We simply do not have enough data on lower-mass stars to know for sure if they are slower than the higher-mass stars in entering this phase (and not all of them do, to the best of our knowledge). Or perhaps the material is currently not being delivered close enough to the star yet for the material to emit strongly. In addition, the sheer {\it volume} of space surrounding the higher-mass stars that is capable of thermally processing grains is larger than in lower-mass stars.

\section{CONCLUSION}

We have presented 3-13 $\mu$m spectra of three stars in the TW Hya Association: TW Hya, HR 4796A, and HD 98800. TW Hya possesses a strong silicate emission band that is generally similar in strength and shape to many other pre-main sequence stars, such as the Herbig Ae/Be stars (HAEBEs) and T Tauri stars (TTs). However, the feature does not show the unambiguous presence of crystalline material, as evidenced by the 11.2 $\mu$m feature, as is observes in some HAEBEs and in some long-period solar system comets. HR 4796A has only a very weak excess emission above photospheric levels, and nothing can be said about the presence of a significant silicate band from these data. HD 98800 is intermediate between these two in terms of the strength of its silicate band, but the data are inadequate to say anything about the mineralogy or degree of crystallinity.

For the object where we have the best data, TW Hya, the lack of obvious crystallinity is not entirely surprising, as it is absent in all of the youngest pre-main sequence stars, and in most but not all of the older ones. It is present in many of the long-period comets of our own solar system, however, and our failure to detect it so far in a younger star of roughly solar mass is intriguing, though not unexpected. We are far from understanding how the crystalline material is produced and incorporated into the grains of our own solar system and in the disk systems of some of the higher-mass stars.

One of the main obstacles preventing more definitive conclusions is the lack of high-quality mid-IR spectra of TTs. Many can be reached using ground-based instruments, but require long (in excess of an hour or two) integration times to achieve the necessary signal-to-noise. A truly comprehensive study will require an instrument like the Space Infrared Telescope Facility (SIRTF).

Support for this work was provided for MLS through NASA's Origins of Solar Systems Program grant NAG5-9475, and the University Research Council and Physics Department of the University of Cincinnati. Support for DKL and RWR was provided by the The Aerospace Corporation's Independent Research and Development program. We would like to thank Ann Mazuk, and Ted Tessensohn for technical support and Bill Golisch and Dave Griep of the IRTF for expert telescope operation. The authors would also like to thank Robert Joseph for being flexible in the scheduling of the IRTF time for this project with our other programs. We also thank Carol Grady for her comments on the manuscript. The IRAS broadband fluxes were obtained via SIMBAD.

\end{document}